%% file: main.tex
\documentclass[conference]{IEEEtran}
\IEEEoverridecommandlockouts
\usepackage{cite}
\usepackage{amsmath,amssymb,amsfonts}
\usepackage{algorithmic}
\usepackage{graphicx}
\usepackage{textcomp}
\usepackage{xcolor}
\usepackage{booktabs}
\usepackage{url}
\usepackage{multirow}
\usepackage{balance}
\usepackage{color, colortbl}
\usepackage{placeins}
\usepackage{makecell}
\usepackage{flushend}
\usepackage{listings}
\usepackage{subcaption}

\newcommand{\ie}{\textit{i}.\textit{e}.,\ }
\newcommand{\eg}{\textit{e}.\textit{g}.,\ }
\newcommand{\cf}{\textit{c}\textit{f}.\ }
\newcommand\1{\textit{(i)}}
\newcommand\2{\textit{(ii)}}
\newcommand\3{\textit{(iii)}}

\usepackage{pifont}
\newcommand{\cmark}{\ding{51}}%
\newcommand{\xmark}{\ding{55}}%

\definecolor{codegreen}{rgb}{0,0.6,0}
\definecolor{codegray}{rgb}{0.5,0.5,0.5}
\definecolor{codepurple}{rgb}{0.58,0,0.82}
\definecolor{backcolour}{rgb}{0.95,0.95,0.92}

\lstdefinestyle{mystyle}{
    backgroundcolor=\color{backcolour},   
    commentstyle=\color{codegreen},
    keywordstyle=\color{magenta},
    numberstyle=\tiny\color{codegray},
    stringstyle=\color{codepurple},
    basicstyle=\ttfamily\footnotesize,
    breakatwhitespace=false,         
    breaklines=true,                 
    captionpos=b,                    
    keepspaces=true,                 
    numbers=left,                    
    numbersep=5pt,                  
    showspaces=false,                
    showstringspaces=false,
    showtabs=false,                  
    tabsize=2
}

\begin{document}

\title{Trust, But Verify: An Empirical Evaluation of AI-Generated Code for SDN Controllers}

\author{\IEEEauthorblockN{Felipe Avencourt Soares, Muriel F. Franco, Eder J. Scheid, Lisandro Z. Granville}
\IEEEauthorblockA{\\Federal University of Rio Grande do Sul (UFRGS), 
Porto Alegre, Brazil\\
Federal University of Health Sciences of Porto Alegre (UFCSPA), Porto Alegre, Brazil\\
\small{\texttt{\{felipe.avencourtsoares, ejscheid, granville\}@inf.ufrgs.br}}\\
\small{\texttt{muriel.franco@ufcspa.edu.br}}
}}

\maketitle

\begin{abstract}
Generative Artificial Intelligence (AI) tools have been used to generate human-like content across multiple domains (\eg sound, image, text, and programming). However, their reliability in terms of correctness and functionality in novel contexts such as programmable networks remains unclear. Hence, this paper presents an empirical evaluation of the source code of a POX controller generated by different AI tools, namely ChatGPT, Copilot, DeepSeek, and BlackBox.ai. To evaluate such a code, three networking tasks of increasing complexity were defined and for each task, zero-shot and few-shot prompting techniques were input to the tools. Next, the output code was tested in emulated network topologies with Mininet and analyzed according to functionality, correctness, and the need for manual fixes. Results show that all evaluated models can produce functional controllers. However, ChatGPT and DeepSeek exhibited higher consistency and code quality, while Copilot and BlackBox.ai required more adjustments. 
\end{abstract}


\input{sections/introduction}
\input{sections/relatedwork}
\input{sections/methodology}
\input{sections/evaluation}
\input{sections/discussion}
\input{sections/conclusion}


\bibliographystyle{IEEEtran}
\bibliography{bib/references.bib}
\footnotesize
All links were visited in October 2025

\end{document}

%% file: sections/introduction.tex
\section{Introduction}

Generative Artificial Intelligence (AI), that is, a type of AI capable of producing human-like content~\cite{codeGeneration}, has been applied in multiple domains, such as text, music, and image generation~\cite{aiApplications}. However, its applications go beyond these cases. One specific example is the automatic generation of source code in various programming languages (\eg C, Python, C++, and Java) based on textual descriptions of program objectives, which, when well-defined~\cite{improvingChatGPTCode}, can produce results that meet the user’s specified requirements. Hence, ``vibe coding", which is practice of developing software through conversations with Large Language Models (LLM) has become popular~\cite{vibeCoding}.

Nevertheless, programs generated by generative AI applications (\eg ChatGPT~\cite{chatgpt}) are not always correct and may pose security concerns~\cite{chatGPTCodeCorrect,vibeCoding}, as they may contain logical errors or performance issues. Therefore, developers cannot blindly trust the output produced by such systems. This problem becomes even more critical when the generated code targets network devices that must ensure security (\eg firewalls) or maintain performance to meet Quality of Service (QoS)~\cite{KHAN2021176} and Service Level Agreement (SLA) policies~\cite{IM19EderAutomaticSLA}.

Although the popularity and use of such tools are increasing rapidly, research on this topic remains limited due to its novelty. Moreover, as computer networks evolve to become increasingly programmable and open~\cite{programmableNetworksOpen}, several efforts have already attempted to automate code generation for these environments~\cite{sdnAIAutomate,IBNAINOMS}, often without properly considering the correctness or reliability of the produced code. Hence, the behavior and trustworthiness of AI-generated code for programmable networks (\eg for Software-Defined Networking - SDN) deserve further investigation.

In this context, this paper analyzes different codes generated by generative AI applications (\eg ChatGPT, CoPilot, DeepSeek, and BlockBox.ai) within the scope of programmable networks (\eg SDN controllers~\cite{sdnCode}) to verify whether they achieve the user-defined objectives, are functionally correct, or present security issues. Finally, it empirically compares and discuss the differences and variations between the codes generated by the selected AI tools.

The remainder of this paper is organized as follows. Section~\ref{sec:rw} reviews the literature on AI-based code generation for both SDN and general-purpose applications. Section~\ref{sec:main} describes the methodology adopted in this study, including the definition of AI tools, prompts, and network topologies used for testing. Next, Section~\ref{sec:experiments} details the experiments conducted with the code generated by the selected tools, while Section~\ref{sec:discussion} discusses the results across different dimensions. Finally, Section~\ref{sec:conclusion} concludes the paper and outlines directions for future work.

%% file: sections/relatedwork.tex
\section{Related Work}
\label{sec:rw}

\begin{table*}[ht]
    \centering
    \caption{Comparision of Related Work}
    \label{tab:programmable_networks}
    \begin{tabular}{@{}cccc c c c@{}}
        \toprule
        \textbf{Reference} & \textbf{Year} & \textbf{Context} & \textbf{Programmable Network} & \textbf{AI Tool/Technique/Model} & \textbf{Code Generation} & \textbf{Evaluation} \\ 
        \midrule
        \cite{networkRepresentations} & 2022 & Networking & No & BERT & No & Study \\ 
        \cite{isYourCodeGenerated} & 2023 & Computing & No & GPTs, CodeGen, StarCoder & Yes & Experiment \\ 
        \cite{learningRepresentations} & 2023 & Computing & No & BERT and GPT-3 & No & Experiment \\ 
        \cite{isChatGPTCapable} & 2023 & Computing & SDN & GPT-3.5 & Yes & Experiment \\ 
        \cite{networkMeetsChatGPT} & 2023 & Networking & SDN & NetLM & No & Study \\ 
        \cite{enhancingNetworkManagement} & 2023 & Networking & No & Bard and GPT-3/4 & Yes & Study \\ 
        \cite{largeLanguageModelsReview} & 2024 & Networking & SDN & Several & No & Discussion \\ 
        \cite{largeLanguageModelsZeroTouch} & 2024 & Networking & No & GPT-4, NetCFG & No & Experiment \\ 
        \cite{LLMIntentProcessing} & 2025 & Networking & SDN & LLM & Yes & Experiment \\ 
        This Work & 2025 & Networking & SDN & ChatGPT, Copilot, DeepSeek, Blackbox.io & Yes & Experiment \\ 
        \bottomrule
    \end{tabular}
\end{table*}

A first step in the research involves the analysis of the state-of-the-art to help identify challenges that require further exploration. Table~\ref{tab:programmable_networks} presents a concise summary relating the two main topics of this work.

The first work \cite{largeLanguageModelsZeroTouch} explores the capability of a Large Language Model (LLM) to autonomously manage a network from scratch through the proposed LLM-NetCFG framework. Another relevant study \cite{isChatGPTCapable}, which had a significant influence on this research, evaluates the ability of ChatGPT-3.5 to solve computing and networking problems. The authors follow a systematic methodology by stating the problem, presenting the expected solution, describing the prompt provided to the LLM, and analyzing the model’s output. However, the paper leaves open the evaluation of other natural language models and notes inaccuracies that have since been corrected due to model evolution.

The review in \cite{largeLanguageModelsReview} provides a broad overview of AI and LLM applications in networking, referencing multiple studies across various subfields since the rise of these technologies. The work in \cite{networkMeetsChatGPT} proposes the implementation of NetLM, a ChatGPT-derived model designed to enable intent-based communication and autonomous network management. The study \cite{learningRepresentations} highlights the role of log analysis for fault prediction and operational support. It employs LLMs such as GPT-3 and BERT, trained on large datasets to improve accuracy and reliability in anomaly detection. 

In \cite{networkRepresentations}, BERT-based learning is adapted to produce semantic numerical representations (embeddings) that enhance model generalization and network device security. The research in \cite{LLMIntentProcessing} examines network automation in 5G environments using LLMs like BERT, focusing on intent interpretation and transformer-based time-series prediction for validation. The framework described in \cite{enhancingNetworkManagement} combines synthesis techniques to automatically generate graph-based code for communication networks. Finally, \cite{isYourCodeGenerated} proposes an automated benchmarking and validation framework for assessing the quality and correctness of LLM-generated code.

It can be concluded from this research that most of the reviewed studies and experiments rely heavily on LLMs, particularly ChatGPT. In general, these models are employed as management or detection tools within their respective applications, with learning derived from large datasets tailored to each context. Among the analyzed works focusing on code analysis generated by LLMs, only \cite{LLMIntentProcessing} is directly related to the programmable networks subcontext and includes code generation. However, it remains limited to the control plane and constrained to a single LLM (\ie BERT).

%% file: sections/methodology.tex
\section{Methodology}
\label{sec:main}

In this section, the AI tols to be analyzed are defined, followed by the definition of the prompts. Next, the network topologies on which these AIs will operate are presented, along with the execution of the respective prompts. Finally, the functionality of the codes generated by the Large Language Models (LLMs) is analyzed.

\subsection{Artificial Intelligence Tools}

A preliminary review of AI reveals a wide range of available options. Given this diversity, specific criteria were established to narrow the scope of analysis, aiming to capture different aspects in terms of both popularity and technical approach. Based on these filters, the following AIs were selected for study: ChatGPT~\cite{chatgpt}, Coplit~\cite{copilot}, DeepSeek~\cite{deepSeekR1}, and BlackBox.ai~\cite{blackbox}.

ChatGPT, developed by OpenAI, is the most popular among those selected and is known for its versatility in solving general-purpose problems. Similar to DeepSeek and BlackBox.ai, it provides a conversational interface in which the user inputs a text description containing the necessary information, referred to here as a \textit{prompt}. The initial interaction stage is crucial, as it guides the solution through keywords. Therefore, a well-crafted prompt results in a more accurate and structured response.

Copilot, developed by Microsoft, also provides a conversational interface similar to the previous models. However, its GitHub Copilot version includes functionality that assists directly within programming environments. This tool analyzes existing code and suggests the subsequent lines to accelerate software development.

DeepSeek, a Chinese company, recently released the DeepSeek-R1 model, which claims to match or even surpass OpenAI’s O1 model in some aspects while offering significantly lower cost. Its training approach differs from that of OpenAI models by emphasizing reasoning trajectories learned from prior model behaviors rather than relying primarily on large-scale web data. This distinction is detailed in the study published by the company itself~\cite{deepSeekR1}.

BlackBox.ai, originally developed by ACG, also features an interactive chat interface. Among the AIs described, it was the first to introduce the functionality of attaching files to task descriptions. Its interface remains minimalist while offering several advanced features to assist developers.

\subsection{Tasks and Prompts}
\label{sec:tarefas-prompts}

To evaluate the models mentioned above (\ie ChatGPT, DeepSeek, Copilot, and BlackBox.ai) three tasks of increasing complexity were formulated, each designed to assess the learning and reasoning capabilities of the models. In addition, two prompt-engineering techniques were selected to compose the commands for each task, zero-shot and few-shot. The defined tasks are as follows:

\begin{itemize}
    \item \textbf{Simple:} Regardless of topology, ensure full packet delivery between all hosts.
    \item \textbf{Intermediate:} According to the topology, block communication between hosts as specified in the prompt.
    \item \textbf{Complex:} Develop a layer-three firewall for an SDN.
\end{itemize}

Initially, the \textit{zero-shot} technique was employed to design the prompts. This approach consists of providing direct and objective instructions to the model, without prior examples or contextual information. The second technique, \textit{few-shot}, incorporates contextual elements and examples to guide the model toward a more precise solution. Hence, reducing ambiguity.

\subsubsection{Zero-Shot}

For the zero-shot technique, the inputs are concise and keyword-oriented to focus the model on the intended outcome. The following prompts were defined:

\begin{itemize}
    \item \textbf{Prompt 1}
    \begin{itemize}
        \item \textit{``Develop, in Python, a POX controller for a software-defined network."}
    \end{itemize}
    \item \textbf{Prompt 2}
    \begin{itemize}
        \item \textit{``Develop, in Python, a POX controller for a software-defined network where it blocks pings between even hosts and odd hosts."}
    \end{itemize}
    \item \textbf{Prompt 3}
    \begin{itemize}
        \item \textit{``Develop, in Python, a layer-three firewall POX controller for a software-defined network."}
    \end{itemize}
\end{itemize}

\subsubsection{Few-shot}

For the few-shot technique, which uses examples or contextual descriptions to refine the response, the following inputs were defined. These are more complex and detailed than the zero-shot prompts, providing richer contextualization of the problem domain.

\begin{itemize}
    \item \textbf{Prompt 1}
    \begin{itemize}
        \item \textit{``I created a simple software-defined network in Python using the Mininet library. In this network, I wanted to control the flow of packets between hosts. To test communication within the network, I needed to start a controller. In this context, make a POX controller that manages packet forwarding between hosts. Additionally, ensure the controller learns MAC addresses. Finally, provide a brief explanation of the implementation."}
    \end{itemize}
    \item \textbf{Prompt 2}
    \begin{itemize}
        \item \textit{``I created a simple software-defined network in Python using the Mininet library. In this network, I wanted to control the flow of packets between hosts. To test communication within the network, I needed to start a controller. In this context, make a POX controller to block pings between even hosts and odd hosts. Additionally, shorten the timeout when the ping block occurs. Finally, provide a brief explanation of the implementation."}
    \end{itemize}
    \item \textbf{Prompt 3}
    \begin{itemize}
        \item \textit{``I created a simple software-defined network in Python using the Mininet library. In this network, I wanted to control the flow of packets between hosts. To test communication within the network, I needed to start a controller. In this context, make a layer-three firewall POX controller to manage the software-defined network with the following functions:"}
        \begin{enumerate}
            \item \textit{Initialize control rules;}
            \item \textit{Manage packet flow based on learned MAC addresses;}
            \item \textit{Consider that the following function is implemented:}
                    
        \begin{lstlisting}[language=Python]
def add_firewall_rules(src_ip, dst_ip, protocol="tcp"):
    rule=f"{src_ip}{dst_ip}{protocol}\n"
    with socket.socket(socket.AF_INET, socket.SOCK_STREAM) as s:
        s.connect(("127.0.0.1", 6633))   # Assuming POX listens on this port
        s.sendall(rule.encode())
    print(f"Rule added: Block {protocol} from {src_ip} to {dst_ip}")
\end{lstlisting}
        \item \textit{Finally, provide a brief explanation of the 
        implementation.}
        \end{enumerate}
    \end{itemize}
\end{itemize}

\subsection{Topologies}
\label{sec:Topologias}

\begin{figure*}[ht]
\centering
  \begin{subfigure}[b]{.245\linewidth}
    \centering
    \includegraphics[width=1\textwidth]{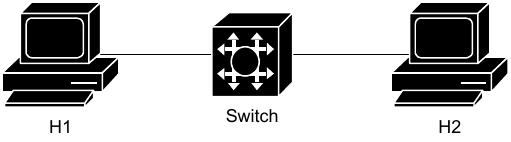}
    \caption{\texttt{Linear}}\label{fig:brasil}
  \end{subfigure}%
  \begin{subfigure}[b]{.245\linewidth}
    \centering
    \includegraphics[width=1\textwidth]{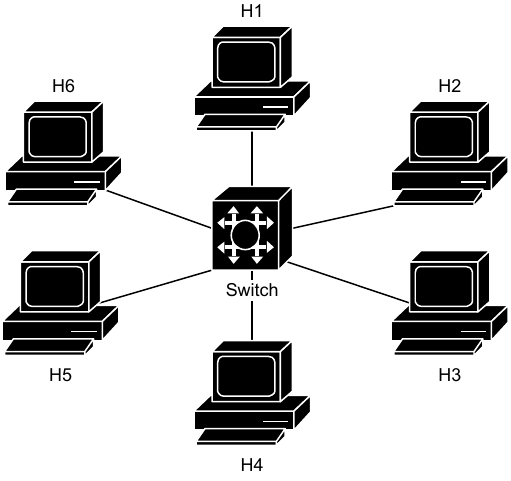}
    \caption{\texttt{Star}}\label{fig:china}
  \end{subfigure}%
  \begin{subfigure}[b]{.245\linewidth}
    \centering
    \includegraphics[width=1\textwidth]{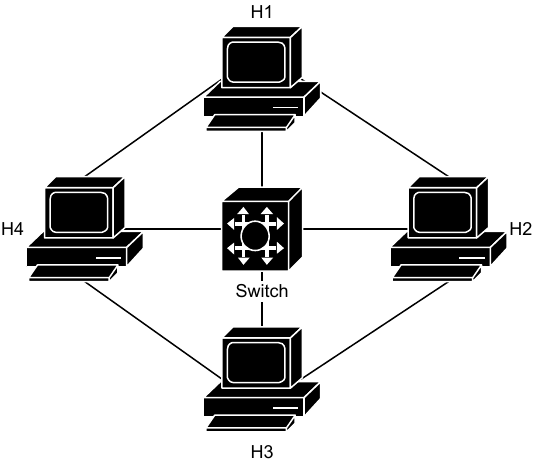}
    \caption{\texttt{Fully Connected}}\label{fig:india}
  \end{subfigure} 
    \begin{subfigure}[b]{.245\linewidth}
    \centering
    \includegraphics[width=1\textwidth]{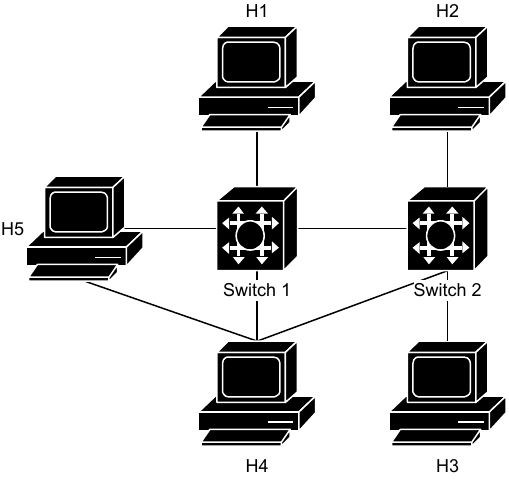}
    \caption{\texttt{Two Switches}}\label{fig:russia}
  \end{subfigure} 
   \caption{Network topologies defined for the tests}
   \label{fig:Topologias}
\end{figure*}

To test the effectiveness of the controllers generated by the AIs, four different network topologies were defined, illustrated in Figure~\ref{fig:Topologias}. These configurations aim to simulate a variety of network scenarios, ranging from simple (\eg linear), medium (\eg star), to complex (\eg, fully connected and two switches).






%% file: sections/evaluation.tex
\section{Experiments}
\label{sec:experiments}


For the experiments, the following methodology was adopted. At the beginning of a new interaction with each AI tool, the input prompts defined in Section~\ref{sec:tarefas-prompts} were provided, corresponding to each prompt-engineering technique and specific task. The generated code was then executed in the POX controller, with network communication simulated using Mininet~\cite{mininet}. If the code executed successfully on the first attempt, it was marked as ``\textit{functional without fixes}"; if additional interaction or manual corrections were required, it was marked as ``\textit{functional with fixes}"; and if it failed after further attempts, it was classified as ``\textit{failure}". All scripts used to generate the network topologies and conduct the tests are publicly available at~\cite{sourceCode}.

\subsection{Task 1 - End-to-end packet delivery among hosts}

\subsubsection{Zero-shot}
\label{sec:Tarefa_1zs}

ChatGPT returned an implementation of the task with an additional feature: learning of MAC addresses, which was not explicitly requested in the prompt. MAC learning associates the source address with the switch port, improving forwarding efficiency. Beyond this optimization, the tool also provided a brief explanation of the generated code and instructions on how to start the controller.

The execution flow is triggered by the packet-in event, which controls packet distribution among network switches. It first performs a packet integrity check, discarding malformed packets and issuing a warning. After validation, the system resolves MAC addresses for proper forwarding, successfully enabling ICMP echo exchanges and meeting the task requirements.

Copilot produced a functional but simplified solution. For each packet event, forwarding is performed via broadcast (all ports), which increases network latency relative to ChatGPT’s approach. Although the control flow is concise, it lacks packet validation and MAC learning; nonetheless, it solves the task.

DeepSeek implemented an SDN controller based on MAC learning similar to ChatGPT’s, but with stronger modularization, separating broadcast, packet sending, and flow-rule installation into dedicated structures.

Like the others, its controller validates packets, learns MAC addresses, and then forwards accordingly. The modular organization improves readability and maintainability while correctly solving the task.

BlackBox.ai also implemented an SDN controller with MAC learning. Its code structure resembles ChatGPT’s solution, while its documentation approach is closer to DeepSeek’s.

The packet-handling mechanism operates as follows: upon a packet-out event, the system checks whether the destination host has been previously learned; if so, it forwards directly on the learned port; otherwise, it broadcasts on all ports. The implementation is functional for the proposed task.

In summary (Table~\ref{tab:resultados-t1-zs}), all selected tools successfully produced a POX controller for SDN. ChatGPT stood out by adding management optimizations; Copilot delivered a simpler solution relying on broadcast; DeepSeek excelled in modular design and readability; and BlackBox.ai-after an update released during this study-balanced strengths seen in ChatGPT and DeepSeek. Even under the zero-shot approach, the AIs demonstrated knowledge of software-based programmable networking.

\begin{table}[h!]
    \centering
    \caption{Task 1 - Results with the Zero-Shot Technique}
    \label{tab:resultados-t1-zs}
    \begin{tabular}{lcccc}
        \toprule
        \textbf{Topology} & \textbf{ChatGPT} & \textbf{Copilot} & \textbf{DeepSeek} & \textbf{BlackBox} \\
        \midrule
        Linear          & \cmark & \cmark & \cmark & \cmark \\
        Star            & \cmark & \cmark & \cmark & \cmark \\
        Fully Connected & \cmark & \cmark & \cmark & \cmark \\
        Two Switches    & \cmark & \cmark & \cmark & \cmark \\
        \bottomrule
    \end{tabular}

    \vspace{2mm}

    \begin{minipage}{\linewidth}
        \centering
        \footnotesize
        \cmark~Functional without fixes;\quad
        \cmark\xmark~Functional with fixes;\quad
        \xmark~Failure
    \end{minipage}
\end{table}

\subsubsection{Few-shot}

ChatGPT subtly improved the code generated in the previous technique. The forwarding routines were refined for clarity and cleanliness. Given the simplicity of the task, basic software-engineering refactoring sufficed to consolidate SDN problem-solving.

Among the tools, Copilot was the only one that had not implemented MAC learning in the zero-shot setting, leaving more room for improvement. Its new response incorporated MAC learning and clarified the algorithm with comments.

As a second-generation LLM with reasoning traces from other systems, DeepSeek’s code is similar to ChatGPT’s. It validates incoming packets before forwarding based on learned MACs. A notable difference is that packet sending was encapsulated in a dedicated function, increasing modularity and readability.

BlackBox.ai attempted to improve modularity relative to its previous version, but the resulting implementation reduced algorithmic clarity. The handler begins with a multicast check and, if not multicast, consults the MAC table for forwarding and learning. Mixing encapsulation logic with packet handling hinders comprehension. A cleaner approach would isolate components (e.g., multicast checks, encapsulation, MAC-table lookups) to improve organization and readability.

Overall, the LLMs demonstrated conceptual mastery of programmable networks and their devices. The additional context in the few-shot approach yielded solutions more aligned with sound software-engineering practices. This qualitative improvement is reflected in the quantitative results in Table~\ref{tab:resultados-t1-fs}.

\begin{table}[h!]
    \centering
    \caption{Task 1 - Results with the Few-Shot Technique}
    \label{tab:resultados-t1-fs}
    \begin{tabular}{lcccc}
        \toprule
        \textbf{Topology} & \textbf{ChatGPT} & \textbf{Copilot} & \textbf{DeepSeek} & \textbf{BlackBox} \\
        \midrule
        Linear          & \cmark & \cmark & \cmark & \cmark \\
        Star            & \cmark & \cmark & \cmark & \cmark \\
        Fully Connected & \cmark & \cmark & \cmark & \cmark \\
        Two Switches    & \cmark & \cmark & \cmark & \cmark \\
        \bottomrule
    \end{tabular}

    \vspace{2mm}

    \begin{minipage}{\linewidth}
        \centering
        \footnotesize
        \cmark~Functional without fixes;\quad
        \cmark\xmark~Functional with fixes;\quad
        \xmark~Failure
    \end{minipage}
\end{table}

\subsection{Task 2 - Blocking pings between hosts}
\label{sec:Tarefa_2}

Two alternative strategies were considered. One would generate a generic controller (as in Task~1) and then request progressive adaptations to meet Task~2 requirements. However, that path would bias the evaluation by introducing preconditions and artificial context. Therefore, we adopted a protocol that starts a fresh dialogue context for this task, presenting the problem independently and without prior understanding.

\subsubsection{Zero-shot}

ChatGPT implemented a helper function to check IP parity. On each packet event, it compares the parity of source and destination IP addresses; if an even-host attempts to communicate with an odd-host (or vice versa), the controller enforces an immediate block. The result meets the requirements with coherent and functional logic.

Copilot created a simple, objective handler that validates IP and ICMP packets and then tests whether to block or forward. The flow is clear and aligns with the task.

DeepSeek provided a direct, functional solution, though with room to improve modularity relative to ChatGPT.

BlackBox.ai followed a similar line to Task~1 but without MAC-learning optimization. It added host-type checks and IPv4/ICMP validation. Finally, it validated parity (even/odd) and enforced the source–destination policy.

In the first prompt, a minor wording error (``between even hosts and odd hosts" instead of “between even hosts with odd hosts”) led to ambiguous interpretation and incorrect solutions from GPT, Copilot, and BlackBox. Only DeepSeek interpreted it correctly. Such issues are common when translating prompts from Portuguese to English. We recommend careful prompt review to minimize ambiguity. After correction, results improved as summarized in Table~\ref{tab:resultados-t2-zs}.

\begin{table}[h!]
    \centering
    \caption{Task 2 - Results with the Zero-Shot Technique}
    \label{tab:resultados-t2-zs}
    \begin{tabular}{lcccc}
        \toprule
        \textbf{Topology} & \textbf{ChatGPT} & \textbf{Copilot} & \textbf{DeepSeek} & \textbf{BlackBox} \\
        \midrule
        Linear          & \cmark & \cmark & \cmark & \cmark \\
        Star            & \cmark & \cmark & \cmark & \cmark \\
        Fully Connected & \cmark & \cmark & \cmark & \cmark \\
        Two Switches    & \cmark & \cmark & \cmark & \cmark \\
        \bottomrule
    \end{tabular}

    \vspace{2mm}

    \begin{minipage}{\linewidth}
        \centering
        \footnotesize
        \cmark~Functional without fixes;\quad
        \cmark\xmark~Functional with fixes;\quad
        \xmark~Failure
    \end{minipage}
\end{table}

\subsubsection{Few-shot}

In this setting, ChatGPT’s initial code mis-mapped IPs in the parity function; a second interaction fixed and improved it. Unlike other tools, it implemented the check as a dedicated \texttt{is\_even} function, reflecting software-engineering learning. The overall logic mirrored prior approaches: ICMP/IPv4 checks followed by host classification (even/odd) to decide to block or to forward.

Copilot’s first answer performed a simple source/destination IP test and blocked when required, but used an incorrect library, causing packet-distribution errors during tests. A second iteration corrected the issue and completed the task.

DeepSeek initially produced a partial solution, requiring a second interaction to fix forwarding. After the fix, the solution was more direct but less structured than before. With the exception of a timeout accelerator initialized early, flow installation and host checks were handled in the packet-event function-undesirable for maintainability.

BlackBox.ai produced a solution similar to its zero-shot version. Improvements included the host parity test and a configurable timeout set at the beginning of the code (akin to constants in other languages).

Although all AIs solved the problem (\cf Table~\ref{tab:resultados-t2-fs}), the few-shot setting revealed greater difficulty in producing well-structured, modular code relative to zero-shot, often requiring more interactions.

\begin{table}[h!]
    \centering
    \caption{Task 2 - Results with the Few-Shot Technique}
    \label{tab:resultados-t2-fs}
    \begin{tabular}{lcccc}
        \toprule
        \textbf{Topology} & \textbf{ChatGPT} & \textbf{Copilot} & \textbf{DeepSeek} & \textbf{BlackBox} \\
        \midrule
        Linear          & \cmark\xmark & \cmark\xmark & \cmark\xmark & \cmark \\
        Star            & \cmark\xmark & \cmark\xmark & \cmark\xmark & \cmark \\
        Fully Connected & \cmark\xmark & \cmark\xmark & \cmark\xmark & \cmark \\
        Two Switches    & \cmark\xmark & \cmark\xmark & \cmark\xmark & \cmark \\
        \bottomrule
    \end{tabular}

    \vspace{2mm}

    \begin{minipage}{\linewidth}
        \centering
        \footnotesize
        \cmark~Functional without fixes;\quad
        \cmark\xmark~Functional with fixes;\quad
        \xmark~Failure
    \end{minipage}
\end{table}

\subsection{Task 3 - Layer-three firewall}

\subsubsection{Zero-shot}

This stage exhibited wide variability due to the unconstrained zero-shot prompt regarding firewall rules. Each tool explored a different solution strategy.

ChatGPT produced a functional IP-based firewall. For each packet event, it checked source/destination IPs against the rule table to block or forward. Forwarding was initially incomplete, but a single-line fix in the send path resolved the issue and the task was completed.

Copilot produced a very similar L3 firewall, but left an explicit \emph{TODO} regarding ARP handling, unlike ChatGPT. ARP maps IP addresses to MAC addresses and must be considered in practical deployments.

After a second interaction to fix a parameter error, DeepSeek delivered a functional firewall with clear separation of initialization, default rules, rule insertion, and packet handling. Forwarding logic followed the familiar IP-based approach.

BlackBox.ai returned code very similar to previous tasks, indicating limited learning about L3 firewalls. The flow consisted of brief initialization with connections and allowed IPs, followed by per-packet IP checks for forwarding. Even after two interactions, the implementation did not complete successfully.

Overall (\cf Table~\ref{tab:resultados-t3-zs}), the LLMs produced responses consistent with the requested functionality under the zero-shot approach. Although more interactions were needed to fix syntax or parameter issues, L3 firewalls were structurally~implemented.

\begin{table}[h!]
    \centering
    \caption{Task 3 - Results with the Zero-Shot Technique}
    \label{tab:resultados-t3-zs}
    \begin{tabular}{lcccc}
        \toprule
        \textbf{Topology} & \textbf{ChatGPT} & \textbf{Copilot} & \textbf{DeepSeek} & \textbf{BlackBox} \\
        \midrule
        Linear          & \cmark\xmark & \cmark\xmark & \cmark\xmark & \xmark \\
        Star            & \cmark\xmark & \cmark\xmark & \cmark\xmark & \xmark \\
        Fully Connected & \cmark\xmark & \cmark\xmark & \cmark\xmark & \xmark \\
        Two Switches    & \cmark\xmark & \cmark\xmark & \cmark\xmark & \xmark \\
        \bottomrule
    \end{tabular}

    \vspace{2mm}

    \begin{minipage}{\linewidth}
        \centering
        \footnotesize
        \cmark~Functional without fixes;\quad
        \cmark\xmark~Functional with fixes;\quad
        \xmark~Failure
    \end{minipage}
\end{table}

\subsubsection{Few-shot}

While the previous approach assessed flexibility, the few-shot methodology here enables deeper exploration via code fragments. With a specific prompt, we evaluated and integrated previously developed code, as detailed in Section~\ref{sec:tarefas-prompts}.

ChatGPT correctly integrated prior code into a working firewall, extending the controller to support local rule insertion, block checking, and packet management with MAC learning.

For rule insertion, it checks whether a rule is already active before installing it. For the block checker, it first validates packet structure, discards malformed packets, identifies the protocol, and then consults the active rule table.

Copilot implemented a firewall with more specialized functionality than in zero-shot while meeting the prompt’s requirements. Notably, it removed per-packet ARP checks in favor of a generalized rule-list mechanism, yielding cleaner code. Like ChatGPT, it supported dynamic local rule insertion.

DeepSeek proposed a distinct approach: an L3 firewall backed by a TCP server. Benefits include \1~real-time rule updates, \2~easier integration with external systems, and \3~higher modularity and scalability. Downsides include higher development complexity and potential security implications. Initialization differs from conventional controllers; other functions (rule definition, insertion, and packet handling) align with methods formalized earlier.

BlackBox.ai implemented an L3 firewall similar to DeepSeek’s but added MAC-learning. Unlike the others, it did not support dynamic local rule insertion. After TCP-server initialization, the algorithm focuses on packet processing and forwarding: \1~integrity validation for MAC learning and \2~IP-specific filtering according to predefined rules. This effectively controls traffic but limits policy update flexibility.

As expected, the higher complexity of this task led to greater variability among the LLMs. ChatGPT and Copilot adopted a consolidated, local controller architecture favoring implementation simplicity and robustness, whereas DeepSeek and BlackBox prioritized dynamic updates, system integration, modularity, and scalability. All solutions satisfied the specified requirements, as summarized in Table~\ref{tab:resultados-t3-fs}.

\begin{table}[h!]
    \centering
    \caption{Task 3 - Results with Few-Shot Technique}
    \label{tab:resultados-t3-fs}
    \begin{tabular}{lcccc}
        \toprule
        \textbf{Topology} & \textbf{ChatGPT} & \textbf{Copilot} & \textbf{DeepSeek} & \textbf{BlackBox} \\
        \midrule
        Linear          & \cmark & \cmark & \cmark & \cmark \\
        Star            & \cmark & \cmark & \cmark & \cmark \\
        Fully Connected & \cmark & \cmark & \cmark & \cmark \\
        Two Switches    & \cmark & \cmark & \cmark & \cmark \\
        \bottomrule
    \end{tabular}
        \vspace{2mm}

    \begin{minipage}{\linewidth}
        \centering
        \footnotesize
        \cmark~Functional without fixes;\quad
        \cmark\xmark~Functional with fixes;\quad
        \xmark~Failure
    \end{minipage}
\end{table}

%% file: sections/discussion.tex
\section{Discussion}
\label{sec:discussion}


In this section, we critically analyze the results of the study. In Section~\ref{sec:Zero-shotVsFew-shot}, we revisit prompt construction to compare the methodological differences between the adopted approaches. Next, in Section~\ref{sec:CompartivoIAs}, we provide a comprehensive comparative analysis of the evaluated tools, highlighting similarities and differences with implementation examples. Finally, in Section~\ref{sec:Debate&Comentários}, we discuss the current technological landscape, characterized by accelerated advances and continuous evolution in SDN and applied AI.

\subsection{Comparison of Approaches: Zero-shot vs Few-shot}
\label{sec:Zero-shotVsFew-shot}

When crafting a first prompt for the task execution, a zero-shot approach can be adopted. As shown in this study, it offers distinct advantages and limitations. This technique, which provides no prior context or training examples, allows greater flexibility and creativity in the responses. However, effective prompt design requires a precise understanding of the problem so that it can be expressed unambiguously and at an adequate level of detail, as in the few-shot approach, thereby aligning the responses with the intended objectives. This perspective is consistent with the reflections of Kevlin Henney, a well-known software engineer and author who has written extensively on programming practice and design clarity. Henney emphasizes that clarity and specificity are central to effective software development~\cite{kevinHenney}.

The importance of precise problem formulation for AI is illustrated by Listings~\ref{code:Copilot_T1_zero-shot} and~\ref{code:Copilot_T1_few-shot}, which show zero-shot and few-shot code, respectively, and exemplify concrete improvements driven by software-engineering principles.

\begin{lstlisting}[language=Python,caption={Copilot-generated code for Task 1 , Zero-shot}, label={code:Copilot_T1_zero-shot}]
def _handle_PacketIn(self, event):
    packet = event.parsed
    log.info("Packet in %s", packet)
    
    msg = of.ofp_flow_mod()
    msg.match = of.ofp_match.from_packet(packet)
    msg.idle_timeout = 10
    msg.hard_timeout = 30
    msg.actions.append(of.ofp_action_output(port=of.OFPP_FLOOD))
    msg.data = event.ofp
    self.connection.send(msg)
\end{lstlisting}

\begin{lstlisting}[language=Python,caption={Copilot-generated code for Task 1 - Few-shot}, label={code:Copilot_T1_few-shot}]
def _handle_PacketIn(self, event):
    packet = event.parsed
    in_port = event.port

    # Learn the source MAC address
    self.mac_to_port[packet.src] = in_port

    # Check if we know the destination MAC address
    if packet.dst in self.mac_to_port:
        out_port = self.mac_to_port[packet.dst]
        log.debug("Sending packet to %s out port %s", packet.dst, out_port)
        msg = of.ofp_packet_out()
        msg.data = event.ofp
        action = of.ofp_action_output(port=out_port)
        msg.actions.append(action)
        self.connection.send(msg)
    else:
        # Flood the packet if the destination MAC is unknown
        log.debug("Flooding packet to %s", packet.dst)
        msg = of.ofp_packet_out()
        msg.data = event.ofp
        action = of.ofp_action_output(port=of.OFPP_FLOOD)
        msg.actions.append(action)
        self.connection.send(msg)
\end{lstlisting}

Although both approaches yield functional solutions, the few-shot technique stands out by producing implementations with greater conceptual refinement, optimized performance, and improved readability. Performance benefits arise from MAC-based forwarding (line~6 of Listing~\ref{code:Copilot_T1_few-shot}), where the switch learns and stores the output port per host.

Regarding readability, and following Fowler’s refactoring principles~\cite{refactoringBook}, code clarity outweighs mere functionality. As observed empirically, algorithms developed with sound software-engineering practices offer significant advantages over ad-hoc implementations, particularly in maintainability and scalability.

\subsection{Comparison Across AI Tools}
\label{sec:CompartivoIAs}

ChatGPT emerged as the most mature and widely used tool among those evaluated. Its responses exhibited consistent technical knowledge of software-controlled programmable networks. In Task~1 under the zero-shot approach, the model proposed an optimization not explicitly requested, indicating contextual inference and generalization. Under few-shot, we observed enhanced responses, incorporating software-engineering best practices and progressive optimizations across consecutive tasks.

Copilot, in turn, required more detailed and contextual prompts to reach the level of learning seen in ChatGPT,especially in software-engineering aspects such as modularity, algorithmic efficiency, and low coupling,relative to ChatGPT and DeepSeek. This limitation reinforces its primary role as a coding assistant within IDEs. Its mechanism relies on real-time static analysis and incremental suggestions for refactoring, bug fixing, and local optimization rather than comprehensive architectural solutions.

DeepSeek-R1, a state-of-the-art language model, outperformed initial expectations, showing consistency and robustness across tasks. Similar to ChatGPT-4, it was able to apply implicit prompt cues to automatically optimize implementations. In few-shot settings, it was effective at assimilating advanced software-engineering concepts, particularly modularization, component cohesion, and algorithmic structure optimization, demonstrating adaptability under limited supervised guidance.

BlackBox.ai, the last tool in this study, initially showed a significant knowledge gap regarding software-based programmable networks, failing to solve many tasks under both approaches. By the end of the study, an important update to the model,motivated by the SDN/AI interplay,improved its outputs for the same prompts, yielding better task completion.

\subsection{Practical Aspects and Additional Remarks}
\label{sec:Debate&Comentários}

LLM-based tools for programmable networks, such as ChatGPT and DeepSeek, bring significant opportunities and challenges to software engineering. These systems evolve rapidly, with frequent updates that can change performance and functionality, requiring continuous adaptation by users. Differences between free and paid tiers, processing capacity, access to advanced models, and longer context windows, can directly impact practical projects. While these tools excel at rapid prototyping and generating modular code, limitations such as response inconsistency, lack of persistent context, and occasional errors reinforce the need for human oversight.

Although few-shot helps reduce inconsistencies through richer contextual descriptions, ambiguity and prompt-induced errors persist. Minor changes in conceptual structure,or simple wording inaccuracies,can lead to interpretive drift, as observed in Task~2. Continuous verification is therefore essential, from prompt design to maintenance of generated code. Even with a well-specified problem, creativity remains key, an idea often echoed in discussions of software craftsmanship~\cite{creativitySE}. Human intervention is thus indispensable for validating and refining AI-produced code, especially in fast-evolving domains such as programmable networks.

Our results show that a network controller can range from basic packet-forwarding rules to more complex systems with advanced features, such as perimeter security management illustrated by the firewall task. Even when correctness is achieved, multiple implementation paths exist, as evidenced by ChatGPT and DeepSeek.

Even when LLMs fully or partially grasp the task, human involvement weighs heavily in the final outcome. Accurate, persistent interpretation from problem conception onward is crucial. Solid initial project scoping,fundamental to effective solutions,helps surface challenges early and enables more optimized, robust final implementations. As seen in Task~1, AIs tend to propose simpler, more direct solutions when not contextualized; small adjustments (\eg enabling MAC learning) can deliver substantial performance improvements.

%% file: sections/conclusion.tex
\section{Summary, Conclusion, and Future Work}
\label{sec:conclusion}

This study analyzed source code generated by different Artificial Intelligence (AI) models in the context of programmable networks. To this end, three prompts were designed, each representing a different level of complexity (simple, intermediate, and complex), and two prompting strategies were employed: \1~zero-shot and \2~few-shot. The objective was to assess the effectiveness and problem-solving capabilities of each language model. For each task, a new chat interaction was initiated following the protocol of the selected prompting technique. From the experimental results, it was observed that the few-shot approach produced more consistent and robust answers, making it a more suitable choice for large-scale~projects.

Overall, two main reasoning patterns emerged among the models: one more consolidated and robust, represented by ChatGPT, and another more innovative and energy-efficient, represented by DeepSeek. Since this research addresses two rapidly evolving areas, programmable networks and AI, it opens multiple avenues for future work. Potential directions include extending the study to P4-based code generation, performing similar analyses with other AI systems (\eg Gemini and custom LLM models), and replicating the experiments in the near future to capture model evolution. Given the fast-paced development of AI technologies, repeating this study even a few months after publication would likely yield distinct and updated results.